\documentclass[11pt]{article}

\usepackage{EACL2023}

\usepackage{times}
\usepackage{latexsym}

\usepackage[T1]{fontenc}

\usepackage[utf8]{inputenc}

\usepackage{microtype}

\usepackage{inconsolata}

\usepackage{graphicx}
\usepackage{subcaption}
\usepackage{caption}
\usepackage{multirow}
\usepackage{booktabs}

\usepackage{cleveref}
\usepackage{enumitem}
\setlist[enumerate]{itemsep=0pt, parsep=0pt, topsep=4pt}
\usepackage[export]{adjustbox}
\usepackage{pgfplots}
\pgfplotsset{compat=1.7}
\usepackage{tikz}
\usepackage{pgf-pie}

\title{DISHONEST: Dissecting misInformation Spread using Homogeneous sOcial NEtworks and Semantic Topic classification}

\author{, , ,  \\
        Pacific Northwest National Laboratory \\ Richland, WA, USA}
        
\author{Caleb Stam,\textsuperscript{1, 2}~ Emily Saldanha,\textsuperscript{1}~ Mahantesh Halappanavar,\textsuperscript{1}~ Anurag Acharya\textsuperscript{1} \\
        \textsuperscript{1}Pacific Northwest National Laboratory, Richland, WA \\
        \textsuperscript{2}University of California San Diego, La Jolla, CA \\
\texttt{\{caleb.stam, emily.grace, mahantesh.halappanavar, anurag.acharya\}@pnnl.gov}}

\begin{document}
\maketitle
\begin{abstract}
The emergence of the COVID-19 pandemic resulted in a significant rise in the spread of misinformation on online platforms such as Twitter. Oftentimes this growth is blamed on the idea of the ``echo chamber.'' However, the behavior said to characterize these echo chambers exists in two dimensions. The first is in a user's social interactions, where they are said to stick with the same clique of like-minded users. The second is in the content of their posts, where they are said to repeatedly espouse homogeneous ideas. In this study, we link the two by using Twitter's network of retweets to study social interactions and topic modeling to study tweet content. In order to measure the diversity of a user’s interactions over time, we develop a novel metric to track the speed at which they travel through the social network. The application of these analysis methods to misinformation-focused data from the pandemic demonstrates correlation between social behavior and tweet content. We believe this correlation supports the common intuition about how antisocial users behave, and further suggests that it holds even in subcommunities already rife with misinformation.
\end{abstract}

\section{Introduction}

The onset of the COVID-19 pandemic was accompanied by an infodemic of online misinformation \cite{Kouzy2020Quant, Cinelli2020Infodemic}. In following years, social media platforms have tried many mitigation strategies to combat anti-vaccine rhetoric and restore trust in scientific expertise \cite{Sundelson2023Infodemic, Calleja2021Infodemic}. However, these strategies were not fully effective. Even in what can now be considered a “post-pandemic” era, extremist communities of misinformation and distrust continue to grow rather than wane \cite{Innes2023PostPandemic}. 

\subsection{Background}

Over the course of the pandemic, we've noted the clear threat misinformation has posed to public health. On social media platforms like Twitter and Facebook, misinformation has been shown to spread faster than the truth \cite{Johnson2020Spread}. Once misinformation has spread online, it seeps into real life. Exposure to online misinformation has been shown to negatively affect individuals intent to vaccinate \cite{Jolley2014PublicHealth, Kricorian2022Vaccine}. More generally, online misinformation threatens public health through encouraging real-life violence and harassment \cite{Shapiro2019threats, Zadrozny2019Harassment}.

Discussion about this spread is largely centered around the echo chamber, a homogeneous community said to form both on social media platforms and in real life. Such chambers are explained via mechanisms which reinforce one's existing opinions, encouraging them to grow more extreme \cite{sunstein1999law}. Studies that model disinformation accordingly have seen success. In particular, the structure of communities on Twitter and Facebook tend to resemble echo chambers.\cite{Cinelli2021EchoChamber}. 

However, the study of the echo chamber has also been criticised. While some accept the echo chamber phenomenon as fact, others criticise it for a lack of empirical evidence or precise definitions. \cite{bruns2021echo}. For instance, there is a lack of consensus on what signals make up the echo. While the conventional idea of an echo chamber suggests individuals are exposed to opinions matching their own, more recent studies have shown an unexpected frequency of cross-cutting exchanges and debates \cite{iyengar2009red, garrett2009echo}. Rather than contradicting the echo chamber phenomenon, this frequency is used to describe a different type of chamber, where constant cross-partisan interaction serves polarizes users further instead of than mediating between them \cite{bright2020echo}.

\subsection{Preliminaries}

Our work approaches the study of antivaccine misinformation on social media from a new angle. We believe that emphasis on the echo chamber as a global structure has obscured the importance of user-level analysis. Instead of modeling a global community structure, we will examine online actors at the individual level. More specifically, we compare two dimensions of a user's online behavior: the speed of their social interactions, and the diversity of the content they post.

First, we are concerned with each user's eagerness or unwillingness to interact with those distant from them in the social network. While the study of spatio-temporal patterns for individual network actors has seen some coverage in the literature \cite{wei2015measuring}, a metric for this quantity has not been definitively established. Therefore, we propose our own, which we refer to as node speed. 

Second, we will analyze the extent to which the content a user posts aligns with misinformation narratives. We leverage natural language processing to achieve this goal. First, we use topic modeling to label posts by their semantic meaning. These topics allow for the computation of several user-level statistics, each of which are compared to node speed in order to draw connections between social and semantic diversity.

\begin{figure*}[t]
    \centering
    \includegraphics[width=1\linewidth]
    {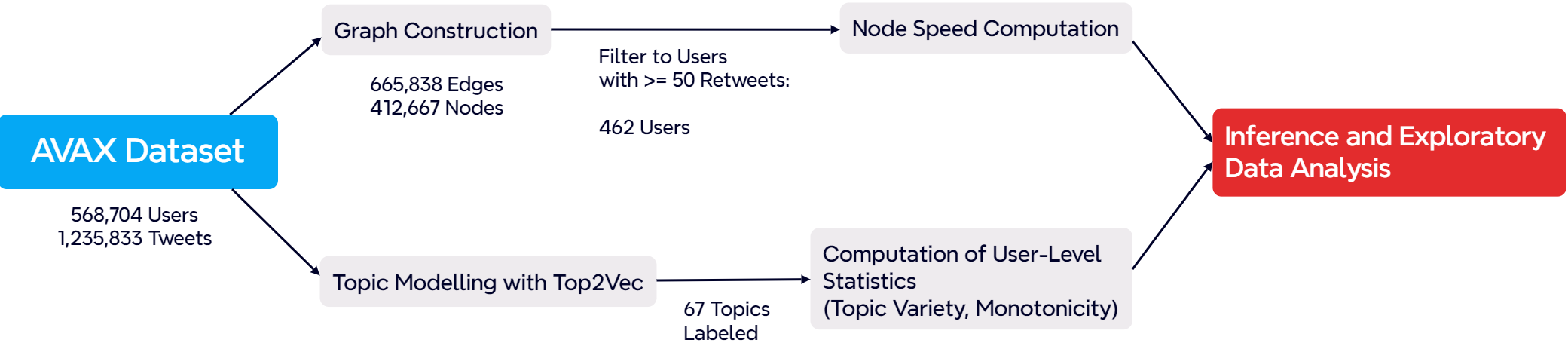}
    \caption{The top-level overview of the methodology of the work presented in this paper}
    \label{fig:methods}
\end{figure*}

\section{Related Work}

The task of analyzing misinformation on Twitter can be approached through the lens of several disciplines. Due to the multimodality of social media data, many computational methods see strong results and provide alternative views of the patterns of information spread and user interactions. In our study, we leverage a combination of semantic topic modeling, dynamic social network analysis, and statistical inference. Further, the problem of misinformation is inherently social. Therefore, methods from the social sciences such as qualitative content analysis see significant use. Existing work in each of these fields has brought advancements in the study of misinformation that lay the foundation for our work.

\subsection{Topic Modeling}

Various topic modeling frameworks, including Latent Dirchlet Allocation \cite{Blei2003LDA}, BERTopic \cite{Grootendorst2022BERTopic}, and Top2Vec \cite{angelov2020top2vec} have been leveraged to analyze social media data from the pandemic. Much of this work has been conducted on activity from the early pandemic. Using topic modeling to comparatively analyze credible and conspiratorial topics from the early pandemic has found that conspiracy-related discussion, while less common than other topics, sees higher engagement than other topics \cite{jiang2021LDAcomptopics}. Furthermore, honing in on misinformation has been demonstrated as an effective and scalable tool to understand the temporal evolution of narratives \cite{marcoux2020LDAmisinfo, Nuzhath2020LDA}. Finally, research using topic modeling from later in the pandemic has leveraged supervised machine learning methods to measure the amount of misinformation within detected topics \cite{Cantini2023BERTopic}.

\subsection{Social Network Analysis}

Other analyses of COVID-19 related misinformation place emphasis on the social network. Approaches inspired by echo chambers have used various community detection algorithms to find clusters of users in networks rife with misinformation, although the structure and nature of these groups varies vastly by domain \cite{AhmedSadri2022CD, Recuero2021Polarization}. Analysis of centrality metrics in these groups have found relationships between various network properties and the likelihood to spread misinformation. Next, the study of information cascades, which can be seen as paths in social networks, has also seen significant work. Rumors and misinformation, both generally and in the context of COVID-19, have been shown to have relatively strong cascades, which could imply more ``sprawl'' in online communities \cite{Vosoughi2018RumorCascades, Solovev2022MisinfoCascades}. Specific research on the structure of information cascades on pandemic-related Twitter networks show that cascades of retweets are large in length but decay faster temporally than other options like quote retweets and replies \cite{DinhParulian2020RTCascades}. Graph neural networks have also been leveraged on social networks to detect misinformation spreaders \cite{Hamid2020GNNs, MaulanaLangguth2023GNNs}. 



\section{Methods}

In this section, we describe our data sources, data processing pipeline, and present an exploratory data analysis along with the analytical pipelines for topic modeling and social network analysis. 

A framework for our methodology is shown in Figure \ref{fig:methods}. Here, we see that our analysis is to be conducted along two main avenues. In the first, we construct a graph representing the social network of Twitter users and analyze it. In the second, we model the topics users are tweeting about with Top2Vec and make inferences about indviduals' tendencies to tweet about varied topics. 

\subsection{Data}

In this study, we leveraged the \textit{Avax} dataset \cite{Muric2021Avax}, which contains 1,235,833 tweets related to vaccine hesitancy from October 2020 to May 2021. This time period is of particular importance to public health, as it contains discussion leading up to and following the widespread distribution of COVID-19 vaccines.

\begin{table}[ht]
    \centering
    \begin{tabular}{lcr}
        \hline
        \textbf{Type} & \textbf{Percentage} \\
        \hline
        Retweet & 59.28\% \\
        Reply & 20.12\% \\
        Original Tweet & 11.93\% \\
        Quote Retweet & 8.67\% \\
        \hline
    \end{tabular}
    \caption{Distribution of tweets by type in the dataset}
    \label{tab:tweet_types_distribution}
    \vspace{-1em}
\end{table}

This dataset was collected using a set of anti-vaccine keywords generated with a snowballing approach. Here, a small list of manually curated antivaccine-related keywords were iteratively expanded by identifying additional co-occuring keywords. This methodology resulted in a set of approximately 60 words and phrases that were queried into the Twitter streaming API to collect tweets. The nature of this data collection process leaves us with tweets mainly focused on anti-vaccine rhetoric and conspiracy theories. 

\begin{figure}[t]
    \centering
    \includegraphics[width=1\linewidth]{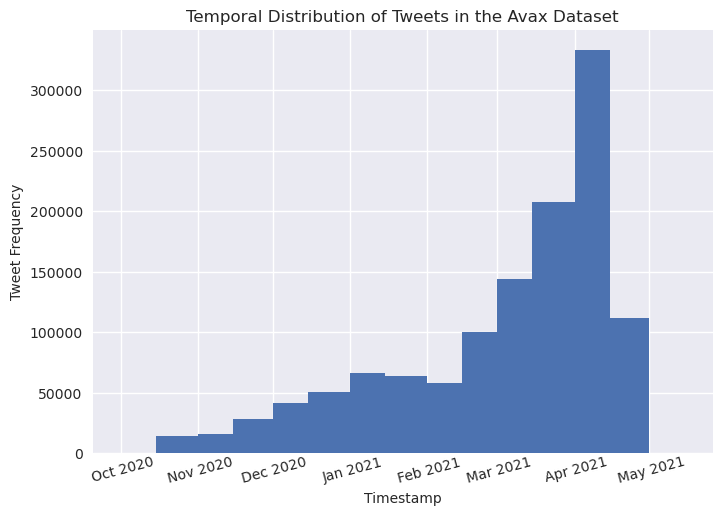}
    \caption{Temporal distribution of the tweets. The tweets cover the period between October 2020 to May 2021.}
    \label{fig:temporal_distribution}
    \vspace{-1em}
\end{figure}

In Figure \ref{fig:temporal_distribution}, we show that the temporal distribution of tweets in the dataset grows over time with a peak in April of 2021. Table \ref{tab:tweet_types_distribution} shows the relative frequencies of each type of tweet in the dataset with the majority of the content being comprised of retweets. We note that this distribution remains approximately constant over time.

\begin{table}[ht]
    \centering
    \begin{tabular}{lr}
        \hline
        \textbf{Metric} & \textbf{Value} \\
        \hline
        Num. Users & 568,704 \\
        Avg. Tweets per User & 2.2 \\
        Max Tweets per User & 1,468 \\
        \hline
    \end{tabular}
    \caption{User statistics for the dataset}
    \label{tab:user_statistics}
    \vspace{-1em}
\end{table}

Since we intend to focus on analysis at the user-level, the distribution of user-level characteristics is of particular importance to us. In Table \ref{tab:user_statistics}, we list relevant descriptive statistics about users in the dataset. Due to the relatively low average number of tweets per user, we next explore the overall distribution of user tweet counts. Table \ref{tab:tweet_threshold} lists the percentage of users in the dataset who have fewer posts than various thresholds. We see that the majority of the users in the dataset only have one recorded tweet, and that a minority (4,821 users) have more than twenty recorded tweets. Thus, despite the large size of the dataset, meaningful user-level analysis needs to be conducted on a small subset of users with sufficient volume for analysis. 


\begin{table}[ht]
    \centering
    \begin{tabular}{cc}
        \hline
        \textbf{Tweet Threshold} & \textbf{Proportion of Users} \\
        \hline
        \(=1\) & 71.47\% \\
        \(\leq2\) & 84.94\% \\
        \(\leq3\) & 90.15\% \\
        \(\leq5\) & 94.53\% \\
        \(\leq10\) & 97.73\% \\
        \(\leq20\) & 99.15\% \\
        \hline
    \end{tabular}
    \caption{Proportion of \textit{Avax} users based on their total number of tweets. Each row shows the proportion of users with a number of tweets equal to or fewer than the given threshold.}
    \label{tab:tweet_threshold}
\end{table}

\subsection{Topic modeling}

\begin{table}[ht]
    \centering
    \begin{tabular}{lccr}
        \toprule
        \textbf{Category} & \textbf{Num. Topics} & \textbf{\% of Tweets} \\
        \midrule
        Policy & 14 & 22.56\% \\
        Conspiracy & 16 & 16.74\% \\
        Science & 5 & 15.96\% \\
        General anti-vax & 6 & 15.78\% \\
        Personal & 8 & 12.04\% \\
        Politics & 5 & 4.56\% \\
        General & 2 & 4.22\% \\
        Meme & 4 & 4.07\% \\
        Rights & 4 & 3.16\% \\
        Negative effect & 3 & 0.90\% \\
        \bottomrule
    \end{tabular}
    \caption{The number of topics in each category and the proportion of tweets in each category. }
    \label{tab:categories}
    \vspace{-1em}
\end{table}

\begin{table*}[ht]
    \centering
    \begin{tabular}{lccl}
        \toprule
        \textbf{Category} & \textbf{Topic} & \textbf{Num. Tweets} & \textbf{Keywords} \\
        \midrule
        \textbf{Conspiracy} & 12 & 20,928 & Depopulation, New World Order, Agenda \\
         & 14 & 49,460 & Depopulation, \#novaccineforme, \#mybodymychoice \\
         & 15 & 21,127 & Depopulation, vaccine, Bill Gates \\
        \textbf{General anti-vax} & 1 & 80,389 & Unvaccinated, Vaccinated, Vaccine, People \\
         & 19 & 59,271 & \#billgatesbioterrorist, vaxxed, \#informedconsent \\
        \textbf{Science} & 5 & 66,358 & Unvaccinated, Vaccinates, Cases, Deaths \\
        \textbf{Policy} & 8 & 50,165 & Unvaccinated, Workers, Working, Health \\
        \textbf{Personal} & 0 & 76,917 & Vaxxed, Gonna,\includegraphics[width=0.035\linewidth,valign=t]{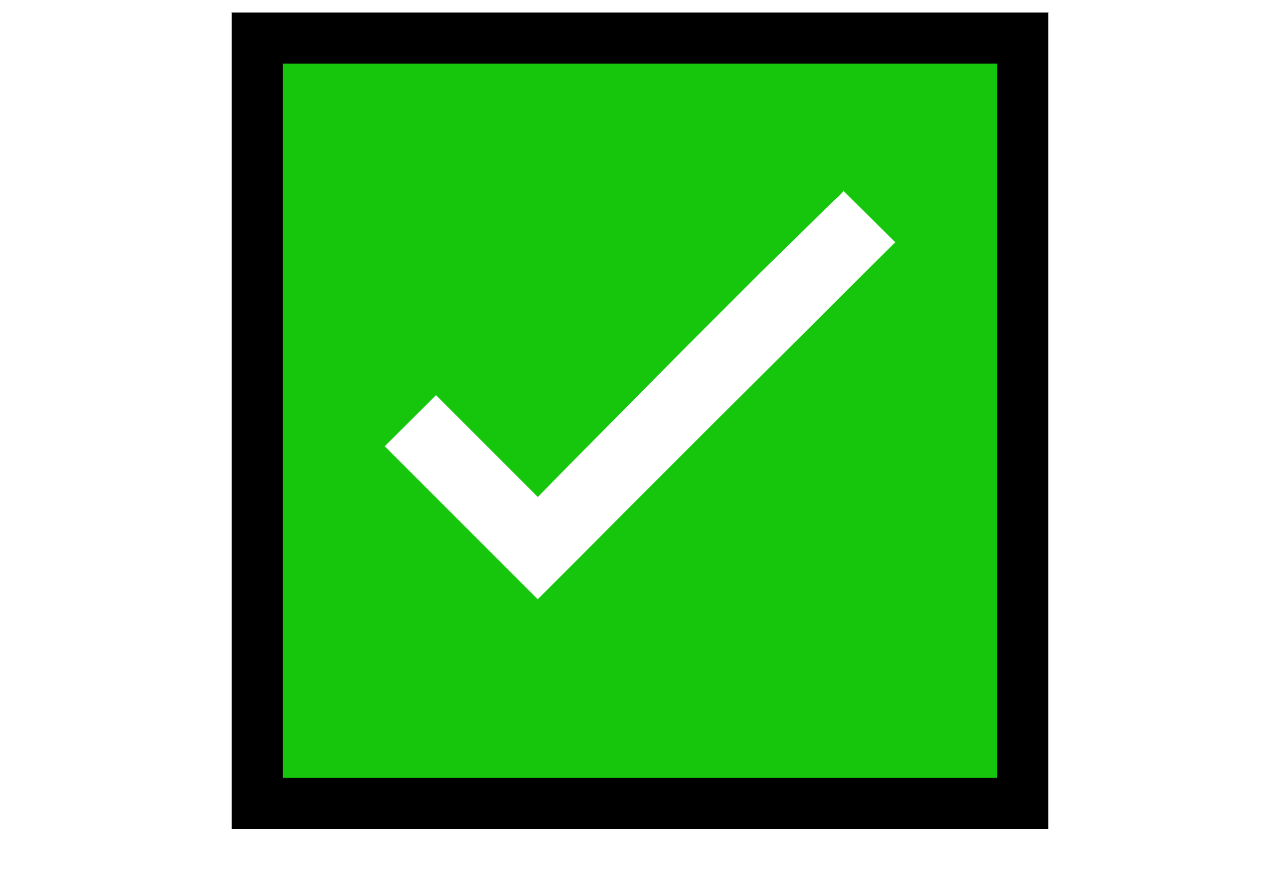} \\
        \bottomrule
    \end{tabular}
    \caption{Top Keywords from Sampled Topics}
    \label{tab:topic_keywords}
\end{table*}

To categorize the most common discussion topics contained in our data, we leverage the Top2Vec topic modeling library \cite{angelov2020top2vec}, which applies hierarchical density-based clustering to the semantic embeddings of a corpus. Embeddings are generated using a sentence embedding model, and these embeddings then undergo dimensionality reduction via UMAP. The HDBSCAN algorithm is used to detect clusters in the reduced dimensionality space.

We cleaned our data before running the topic modeling by eliminating duplicate results. To do so, we dropped all retweets and removed tweets with case-insensitive string matches after removing mentions, whitespaces, and punctuation. This filtering brought our total number of unique tweets to 404,153. We ran the topic model using the universal-sentence-encoder pretrained embedding and clustering parameters {\em min\_cluster\_size = 300} and {\em cluster\_selection\_method = ``leaf''}. This resulted in a total of 67 topics. These topics were further categorized into one of ten categories through manual content analysis. The relative frequency and number of topics contained in each category can be seen in Figure \ref{tab:categories}. The size of a selected set of these topics and their prominent keywords are shown in Figure \ref{tab:topic_keywords}.


\subsection{Graph Construction}

The \textit{Avax} dataset does not include information on which Twitter users follow each other, nor on the relations between users established by liking another's tweets. This leaves us with four interactions we could choose to use as edges: retweets, quote retweets, replies, and mentions. In this study, we choose to use retweets.  We do so in order to assure that edges in our social network connect users of similar sentiment. While the other three possible types of edges (quote retweets, replies, and mentions) are all accompanied by text, retweets are the only edge lacking a text component. From qualitative sampling of the three text-based edges, we notice that a large proportion of these edges represent disagreement between Twitter users. On the other hand, we can typically assume that a retweet of another tweet, without accompanied text, implies agreed sentiment between the user making the retweet and the creator of the original tweet. Because, in this study, we are interested in the narratives that users identify with, not just the ones we interact with, we filter our edges to only retweets. Lastly, because retweets are the most common type of tweet in the dataset, we do not believe that omitting the other three options leads to significant loss of data.

We use the following methodology to construct the social network graph. First, we collect all retweets from the \textit{Avax} dataset. If a retweet exists, in either direction, between any two users, we add an edge between the two users. The representation of the graph in python is created using the NetworkX library \cite{hagberg2020networkx}. This results in an unweighted, undirected social network. We choose not to weight our graph in order to support the computation of the metric described in the next section. 

\begin{figure}
    \centering
    \includegraphics[width=1\linewidth]{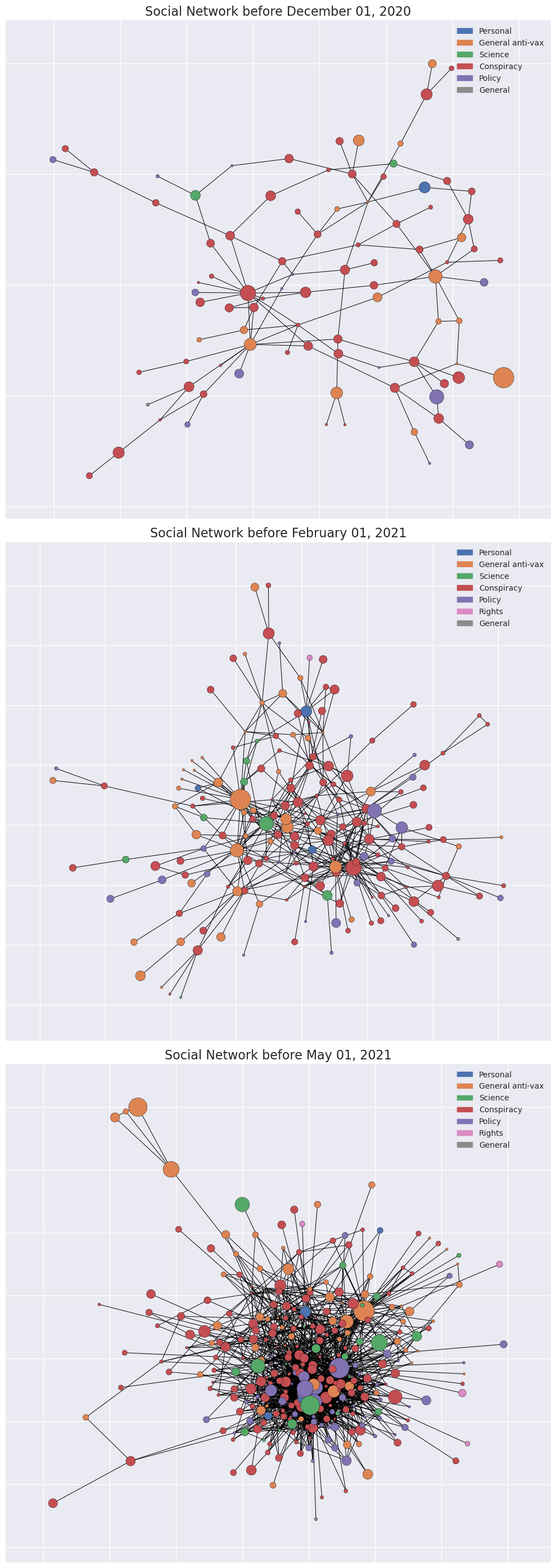}
    \caption{A Growing Social Network: The largest connected component in \textit{Avax} at three different timestamps. Filtered to users with high total tweet counts for interpretability; node size represents the total number of likes received and node color indicates the most tweeted topic category.}
    \label{fig:social_network}
\end{figure}

Since we are concerned with how users move through the social network over time, we take multiple temporal snapshots of this social network. We divide the full span into bins of two weeks each and generate the cumulative set of retweet interactions until the start of each bin. Three visualizations of the graph taken at different timestamps are shown in Figure \ref{fig:social_network}.

\subsection{Node Speed}
In our analysis of the social network, we seek a quantitative method, for each user, to measure the rate at which they make diverse connections in the social network. Such a measure must incorporate both spatial and temporal aspects of the graph. Users who are constantly establish new connections with other users exhibit diversity in the sense that they expose themselves to more users. However, intuition suggests that an interaction joining a user with a faraway region part of the social network is represents more diversity than one made within their existing clique. 

In this sense, we can interpret diversity as speed. A diverse, or fast user, will constantly gain more edges, and those edges will result in them visiting or traveling to different corners of the network. On the other hand, a slow user will stagnate, either failing to establish new connections or only connecting with those users already in their existing social circle.

We establish node speed as a computational framework that aligns with this intuition. Let \(S(u,t)\) be the speed of a node $u$ at snapshot \(t\). We calculate \(S(u,t)\) as:
\begin{equation}
S(u,t)=\frac{\sum_{i=1}^{|V|} d_{t-1}(u,v_i)}{n},
\end{equation}
where \(d_{t-1} (u,v)\) is the shortest path length between nodes $u$ and $v$ at time $t - 1$ (the snapshot prior to $t$). 
The set \(V=(v_1,v_2,...,v_n)\) is the set of nodes \(v_i\) that the node \(u\) creates a new link with during snapshot \(t\), and \(n\) is the total number of new links the user makes in a new snapshot. In the case of retweet data, this includes retweets users that node \(u\) may already have edges with, so \(n\geq|V|\). We choose to normalize with $n$ rather than $|V|$ so that users who repeatedly connect with the same accounts are considered slower than those who primarily interact with accounts that are new to them.

The speed of a node at a particular snapshot can be defined as the average shortest path length between that node and all nodes it established new connections with during the previous timestamp. We expect nodes whose new connections are near their existing communities to have lower speeds, while nodes creating new connections from a faraway part of a social network will have higher speeds. The final value of the node speed is computed as the average of \(S(u,t)\) over all snapshots with at least one retweet.

This methodology requires the handling of a few common edge cases. The value of \(\sum_{i=1}^{n} d_{t-1}(u,v_i)\) cannot be computed in two cases: $(i)$ When node \(v_i\) does not exist in the social network at time \(t-1\), and  
$(ii)$ when \(v_i\) exists at \(t-1\), but was disconnected from \(u\).

We determined that the best solution to work around this problem was to treat such nodes, in either case, as having a shortest path length of one. We did so because of the network structure of our data. In practice, we found that users representing case two were almost always in a connected component of exactly two users. Therefore, instead of thinking of them as being a member of a separate community, we consider them as ``floating'' users, who, like the users representing case 1, do not bear any strong connections to any part of the social network. With this in mind, we consider these users as being close to \(u\) as soon as they connect with it. This motivates our decision to give these ``floating'' users a shortest path length of one. In practice, the choice of how to handle these edge cases had little effect on the relative rank of users' speed.

\section{Results}
We next apply these methods to analyze  the \textit{Avax} dataset. In order to preserve the quality of our node speed metric, and to ensure confidence in aggregated distributions of users' topics, we consider the subset of users in \textit{Avax} who have made at least 50 retweets. This gives us a total of 462 users to analyze. 




\begin{figure}
    \centering

    \begin{subfigure}{\linewidth}
        \centering
        \includegraphics[width=\linewidth]{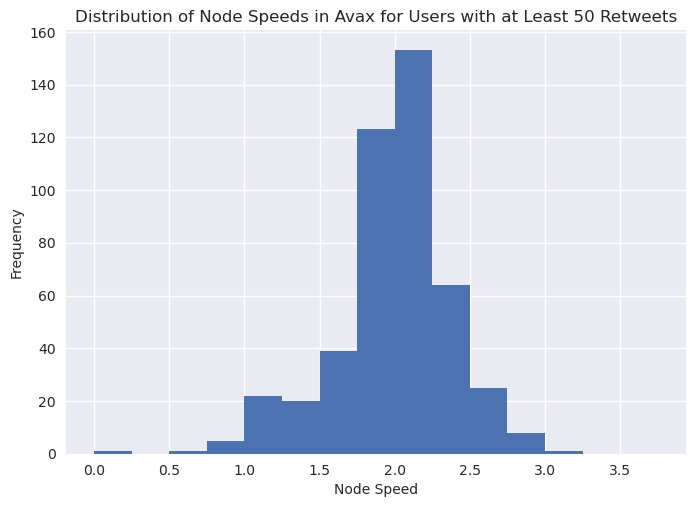}
        \caption{Distribution of Node Speeds of frequently retweeted users (num of retweets $\geq 50$)}
        \label{fig:node_speed_dist}
    \end{subfigure}
    
    \begin{subfigure}{\linewidth}
        \vspace{1em}
        \centering
        \includegraphics[width=\linewidth]{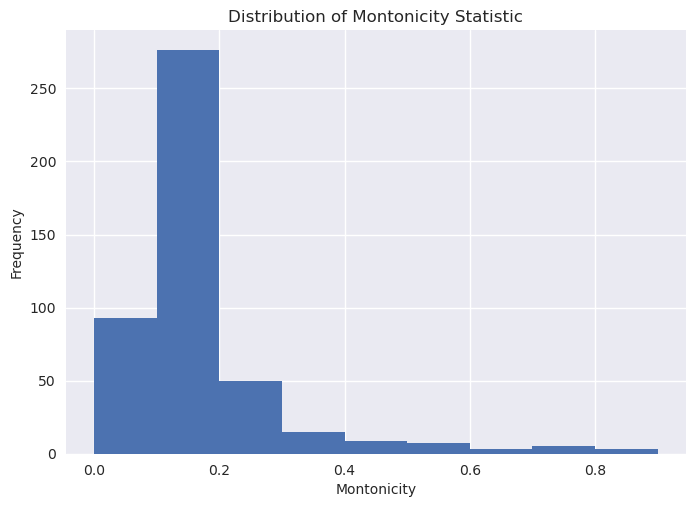}
        \caption{Distribution of the Monotonicity statistic for each user in the dataset}
        \label{fig:monotone_hist}
    \end{subfigure}

    \begin{subfigure}{\linewidth}
        \centering
        \includegraphics[width=\linewidth]{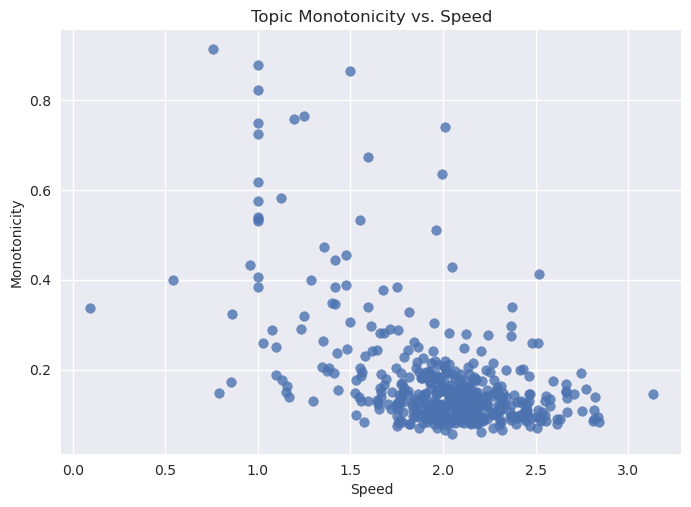}
        \caption{A scatter plot of Node Speed and Topic Monotonicity}
        \label{fig:speed_vs_monotone}
    \end{subfigure}

    \caption{The node speed and monotonicity distribution, a scatter plot of speed and monotonicity}
    \label{fig:combined_figure}
\end{figure}

\subsection{Node Speed in Isolation}

As shown in Figure \ref{fig:node_speed_dist}, the distribution of node speeds is moderately skewed left, with a mean of 1.987 and skewness -0.88.  Thus, we expect to see a larger population of nodes that are slower than the average user in \textit{Avax} than the population that are faster than the average user. There was no apparent correlation between a user's total number of retweets and their speed.

Past studies indicate that the tendency Twitter users involved in COVID-19 discourse to mingle is skewed toward asociality \cite{Sacco2021Pessimism}. Our qualitative analysis seems to support this outlook. Outliers with low speed are truly extreme cases. For example, for one user with speed 1.25, 123 out of 129 of their recorded tweets were retweets from a single user. Taking a closer look, the user was retweeting an account that regularly posted vaccination statistics for the UK. While this is an interesting result to see in our dataset, it doesn't necessarily align with our intuition. Rather, we observe that users with strange social behavior also have strange behavior with respect to content.

Once we cross the threshold of about 1.5, the users interactions are more typical. A particular user with a speed of around 1.65, for instance, appears to be our prototypical low-social-diversity user. While they interact with a more varied set of users than the previous user, their interactions are still relatively concentrated to just a few accounts. 

Lastly, we see that our fastest users have a more diverse spread of interactions. For instance, consider a user with speed 2.22. If we consider the top three accounts that this user retweeted, we notice they make up 30\% of the user's total retweets. Calculating the same metric for the user with speed 1.65 shows that 50\% of their tweets are concentrated in three accounts. This is expected, as a wider variety of retweeted should generally lead to the inclusion of longer paths in the node speed computation. Next, we move towards comparing network speed with semantic topic labels.

\subsection{Speed and Change in Topic}

\begin{figure}
    \vspace{1em}
    \centering
    \includegraphics[width=\linewidth]{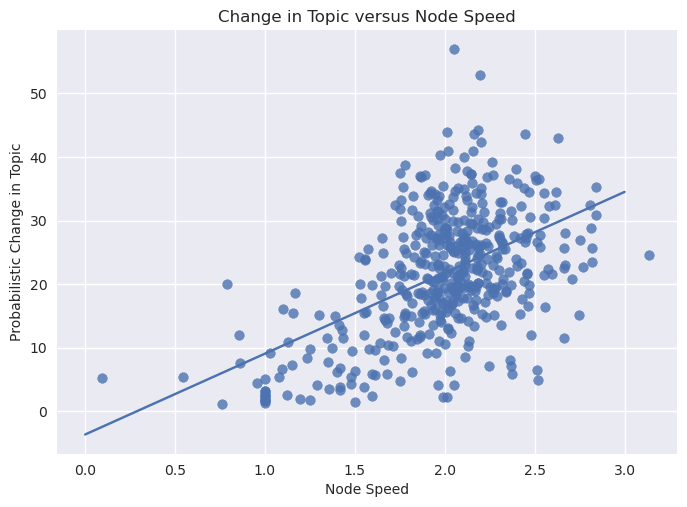}
    \caption{Regression Line between Node Speed and Topic Variation. \(R^2 = 0.27, p=2.54*10^{-33}\)}
    \vspace{1em}
    \label{fig:sup_regression}
    \vspace{-1em}
\end{figure}

One way to link node speed to our topic labels is by measuring the change in a user's choice of topics over time. We use a simple probabilistic approach to quantify this. For each new tweet a user posts, we check the proportion of the user's prior tweets belonging to the same topic. By averaging this result over all of the users' tweets and taking the inverse, we quantify how surprising the topic of each new tweet is for the given user, providing a measure of the variation over time of a user's categorical distribution of tweets. 

Plotting this measure of change in topic against node speed, we find that there is a relatively low, but highly significant positive linear correlation. The regression line, along with its coefficient of determination and p-value for non-correlation are shown in \ref{fig:sup_regression}. This result is expected. Naturally, we would expect slow users to tweet about the same topics repeatedly. On the other hand, fast users should tweet more diversely, covering different topics. However, because of the natural variation inherent to the actions of Twitter users, there is still significant variability in the observed relationship. 

\subsection{Topic Monotonicity in Slow Users}

\begin{figure}

\end{figure}
Beyond looking at the overall distribution of a user's choice in topic, we also see interesting results when examining users who tend to tweet primarily about a single topic. For each user, we compute the maximum proportion of their tweets belonging to a single topic, and consider this to be their monotonicity. The distribution of the monotonicity statistic is shown in Figure \ref{fig:monotone_hist}. Here, we see a distribution that is strongly skewed to the right. As seen in \ref{fig:speed_vs_monotone}, this skew and the high noise makes it difficult to establish a definite linear correlation. However, we do notice a relationship between the extreme values in each statistic.

It appears that users that lie on the right tail of the monotonicity statistic also tend to be on the left tail of the node speed distribution. More specifically, consider the users with monotonicity greater than 0.3. In this data, this value is approximately both one standard deviation above the mean and the 90th percentile of monotonicity. We see in \ref{fig:monotone_hist} that these users are spread out across the distribution's right tail. Out of these users, around 70\% have a node speed less than 1.5, being abnormally slow users (refer to Figure \ref{fig:node_speed_dist}).

\section{Discussion}

The patterns we see in our results support the idea that sections of Twitter with widespread with misinformation are generally antisocial. Despite this, we see that even in these sections users with more diverse social interactions tweet about more diverse topics. This suggests that topics may travel in groups, rather than individually. This fact complicates the challenge of suppressing misinformation. Many existing approaches to the mitigation of misinformation tackle the problem at an individual level, focusing on one topic at a time \cite{halappanavar2021analysis}. Our results motivate a more holistic approach. 

Another interesting result was the overlap of extreme users in speed and monotonicity. Our findings seem to suggest that a complete focus on a single topic implies a lack of participation in the wider social community at all. From our qualitative analysis, we believe these users are best seen as outliers. While they may not fit into our broader framework, they demonstrate the ability to completely isolate even on social platforms. 

Our methodology in this study is also widely generalizable to other applications in the computational social sciences. For any dynamic network incorporating text data, an analysis following this framework can help to identify overarching patterns in the social and semantic diversity of network agents. Even without text data, our node speed metric could be applied in a variety of dynamic social networks beyond social media platforms.

Furthermore, we believe that our work could see meaningful application in misinformation-aware recommender systems. A primary principle of recommender systems is that of homophily, where like-minded users tend to be interested in the same content. However, this can lead to the creation of a filter bubble effect, a cousin of the echo chamber that lowers the diversity of the content users are recommended \cite{Fernandez2020Recommender}. While the existence echo chamber effect in social networks is debated, filter bubbles in recommender systems represent a more obvious threat. Reducing it and its tendency to polarize is one of the primary goals of research in misinformation and harm-aware recommender systems \cite{Tommasel2021Recommender}. 

Our work could improve recommender system algorithms by providing context about a user's likelihood to respond well to more diverse recommendations. For instance, diverse recommendations could be pushed towards users who lie above the regression line in Figure \ref{fig:sup_regression}. These users have high topic variety but low social variety, suggesting that they may be more responsive to diverse perspectives, but are not exposed to them due to their lack of diversity in social connections.

\section{Limitations \& Future Work}

We believe the primary limitation of our work is dataset incompleteness. Amendments to the academic access policy for Twitter's API in 2023 included barriers that prevented us from being able to extract a high volume of tweets. Consequently, we were unable to access each user's historical record of tweets, and instead could only consider those that are included the Avax streaming dataset. While a complete dataset would improve confidence in our results, we believe that the subset of tweets in Avax is a sufficient sample for our purposes.

The second limitation of our work is the confidence in the node speed metric. We developed this metric to get a per-user measure of ``speed'' in a social network. We believe that for this particular use case, the metric was sufficiently validated by the results of its application to the \textit{Avax} data, the simplicity of its calculation, and its similarity to existing metrics \cite{wei2015measuring}. However, we believe a deeper investigation into the properties of node speed, such as its variance or sensitivity to noise, is necessary for more granular data.

Lastly, we believe that further work focused on incorporating the notion of influence could yield powerful results. We have seen that users tweeting about a variety of topics tend to migrate faster throughout the social network. If this speed also represents influence, then this would suggest that misinformation narratives spread as a group rather than independently.


\section*{Acknowledgements}
The research described in this paper was conducted under the Laboratory Directed Research and Development Program at Pacific Northwest National Laboratory, which is operated by Battelle Memorial Institute for the U.S. Department of Energy under contract DE-AC05-76RLO1830. Any opinions, findings, and conclusions or recommendations expressed in this material are those of the author(s) and do not necessarily reflect the views of the United States Government or any agency thereof. This article has been cleared by PNNL for public release as PNNL-SA-193475.

\label{sec:conclusions}

\bibliography{anthology,custom}
\bibliographystyle{acl_natbib}

\appendix



\end{document}